\documentclass[a4paper,fleqn,usenatbib]{mnras}

\usepackage{newtxtext,newtxmath}

\usepackage[T1]{fontenc}
\usepackage{ae,aecompl}

\usepackage{graphicx}	
\usepackage{amsmath}	
\usepackage{rotating}

\newcommand{\Eexc}{$E_{\rm exc}$}
\newcommand{\Teff}{T_{\rm eff}}
\newcommand{\logg}{\rm log~ g}
\newcommand{\kms}{km\,s$^{-1}$}
\newcommand{\eps}[1]{\log\varepsilon_{\rm #1}}
\newcommand{\eu}[5]{\mbox{$#1\,^#2{\rm #3}^{#4}_{\rm #5}$}}
\def\ione{\,{\sc i}}
\def\ii{\,{\sc ii}}
\def\iii{\,{\sc iii}}
\def\iv{\,{\sc iv}}

\title[Non-LTE effects for Sc\ii-\iii\ in A-B stars]{Influence of departures from LTE on determinations of the scandium abundances in A-B type stars}

\author[L. Mashonkina]{
Lyudmila Mashonkina,$^{1}$\thanks{E-mail: lima@inasan.ru}
\\
$^{1}$Institute of Astronomy of the Russian Academy of Sciences, Pyatnitskaya st. 48, 119017, Moscow, Russia
}

\date{Accepted XXX. Received YYY; in original form ZZZ}

\pubyear{2023}

\begin{document}
\label{firstpage}
\pagerange{\pageref{firstpage}--\pageref{lastpage}}
\maketitle

\begin{abstract}
We developed a comprehensive model atom of Sc\ii -Sc\iii. Abundances of scandium for a sample of eight unevolved A9–B3 type stars with well-determined atmospheric parameters were determined based on the non-local thermodynamical equilibrium (NLTE) line formation for Sc\ii -Sc\iii\ and high-resolution observed spectra. For the Sc\ii\ lines, the abundance differences between NLTE and LTE  grow rapidly with increasing effective temperature ($\Teff$), from slightly negative at $\Teff$ = 7250~K to positive ones of up to 0.6~dex at $\Teff$ = 10400~K. For Sc\iii\ in $\iota$~Her, NLTE reduces the line-to-line scatter substantially compared to the LTE case. The NLTE abundances of Sc in our five superficially normal stars are consistent within the error bars with the solar system Sc abundance, while the LTE abundances of the late B-type stars are greatly subsolar. NLTE reduces, but does not remove a deficiency of Sc in the Am stars HD~72660 and Sirius. Based on our own and the literature data, the Ca/Sc abundance ratios of the sample of 16 Am stars were found to be close together, with [Ca/Sc] = 0.6-0.7. We propose the Ca/Sc abundance ratio, but not abundances of individual Ca and Sc elements to be used for classifying a star as Am and for testing the diffusion models.
We provide the NLTE abundance corrections for ten lines of Sc\ii\ in a grid of model atmospheres appropriate for A to late B-type stars.
\end{abstract}




\section{Introduction}

Scandium is one of the key chemical elements in classifying an A-type star as metallic-line (Am) star. \citet{1970PASP...82..781C} justified the following definition: "The Am phenomenon is present in stars that have an apparent surface underabundance of Ca (and/or Sc) and/or an apparent overabundance of the Fe group and heavier elements". He stressed further: "Common to all Am stars is a difference between the Ca (Sc) and other metallic lines". The A-B stars with close-to-solar element abundance pattern (superficially normal stars) and Am stars are being extensively studied \citep[for example][]{1986A&A...163..333H,1990A&A...240..331L,1993AA...276..142H,1999A&A...351..247V,2000MNRAS.316..514A,2007AA...476..911F,2010A&A...523A..71G,2014A&A...562A..84R,2015PASP..127...58A}, however, no additional useful chemical fingerprints of the Am phenomenon have been suggested. In the absense of notable magnetic fields in Am stars, atomic diffusion resulting from the competitive action of gravitational settling and radiative accelerations was proposed by \citet{1970ApJ...160..641M,1980AJ.....85..589M} to explain an origin of chemical peculiarities in the surface layers of Am stars. The diffusion models were recently treated by \citet{2022A&A...668A...6H} in order to reproduce the observed surface abundances of Sc in Am stars.

Classical abundance analysis under the assumption of local thermodynamic equilibrium (LTE) still dominates studies of A-type stars and can result in a wrong classification of Am stars. One educative example is HD~209459 (21~Peg). \citet{2009AA...503..945F} found the LTE abundances of Ti, V, Mn, Fe, Co, and Ni in this star to be consistent within the error bars with the solar abundances, but lower abundances of Ca and Sc, by 0.2~dex and 0.4~dex, respectively. In addition, 21~Peg reveals an enhancement in barium, with [Ba/H] = 0.68. The LTE abundances favour a classification of 21~Peg as a hot Am star. Only close-to-solar abundances of the Fe group elements prevented \citet{2009AA...503..945F} from such a wrong conclusion.
The LTE deficiency of Ca in 21~Peg was removed in the calculations based on the non-local thermodynamic equilibrium (non-LTE, NLTE) line formation \citep{2018MNRAS.477.3343S}. \citet{2020MNRAS.499.3706M} have shown that the NLTE abundance pattern of 21~Peg in the He to Fe range is compatible with the solar one, and thus 21~Peg is a superficially normal star. Though scandium was treated under the LTE assumption and \citet{2020MNRAS.499.3706M} speculated about a positive NLTE abundance correction at the level of 0.5~dex. As for the heavy elements, NLTE leads to even higher abundance of barium, [Ba/H] = 0.98, and enhancements in the other heavy elements, [Sr/H] = 0.59 and [Zr/H] = 0.35. Enhancements in the heavy elements growing towards the higher effective temperature ($\Teff$) seem to be a characteristic of not only Am, but also superficially normal A-type stars, though in the latter stars, enhancements are of less magnitude \citep{2020MNRAS.499.3706M}.

The NLTE calculations were not performed yet for scandium in the $\Teff >$ 9000~K range. In the atmospheres with $\Teff >$ 9000~K, scandium is mostly represented by the doubly ionized atoms, Sc\iii, while the Sc abundances are derived from lines of Sc\ii. 
Number density of Sc\ii\ in such atmospheres is sensitive to the intensity of ionizing ultraviolet (UV) radiation and subject to the departures from LTE.  
The model atom has to include levels and transitions of Sc\ii\ and Sc\iii.

For the cooler stars, such as the Sun and G-K type stars of different metallicities, NLTE analyses of the Sc\ii\ lines 
are made with the model atom of Sc\ione -Sc\ii\ first treated by \citet{Zhang2008_sc} and with the Sc\ii\ model atom of \citet{nlte_sc2}. The latter model atom was also applied to derive the NLTE abundance of scandium in an Am star HD~180347, which has $\Teff$ = 7740~K \citep{2023MNRAS.524.1044T}. In the atmosphere of HD~180347, Sc\ii\ is the majority species, and its statistical equilibrium is correctly calculated with the model atom that does not include excited levels of Sc\iii. 

This study aims to develop a NLTE method for analysis of the Sc\ii\ and Sc\iii\ lines in spectra of A- to mid B-type stars. We continue the theoretical research of the NLTE line formation in the atmospheres of A-B type stars that are presented in a series of papers by \citet{sitnova_o,sitnova_ti,2018MNRAS.477.3343S,2022MNRAS.515.1510S} for O\ione, Ca\ione -\ii, Ti\ione -\ii, and Zn\ione -\ii, \citet{2016MNRAS.462.1123A,2018ApJ...866..153A,2020ApJ...896...59A} for C\ione -\ii, Mg\ione -\ii, and Ne\ione, \citet{2020MNRAS.493.6095M} for Si\ione -\ii -\iii, and \citet{2020MNRAS.499.3706M} for Na\ione, Sr\ii, Zr\ii -\iii, and Ba\ii.


The paper is organised as follows. A comprehensive model atom of scandium is first presented in Sect.~\ref{sect:model_atom}. 
Stellar sample, spectral observations, and atmospheric parameters are described briefly in Sect.~\ref{sect:obs}. Section~\ref{sect:nlte} shows the obtained NLTE effects on level populations and spectral lines of Sc\ii\ and Sc\iii. The LTE and NLTE abundances of Sc are determined and discussed in Sect.~\ref{sect:abund}. In Sect.~\ref{sect:nlte_corr}, we calculate the NLTE abundance corrections for lines of Sc\ii\ in a grid of model atmospheres appropriate for A-type stars. 
The conclusions are summarised in Sect.~\ref{sect:Conclusions}.

\section{Model atom of scandium}\label{sect:model_atom}

We take the model atom of Sc\ii\ developed by \citet[][hereafter, MR2022]{nlte_sc2} as the basis and  update it by including energy levels and transitions of Sc\iii. We describe briefly the used atomic data.

\begin{figure*}
	\includegraphics[width=\linewidth]{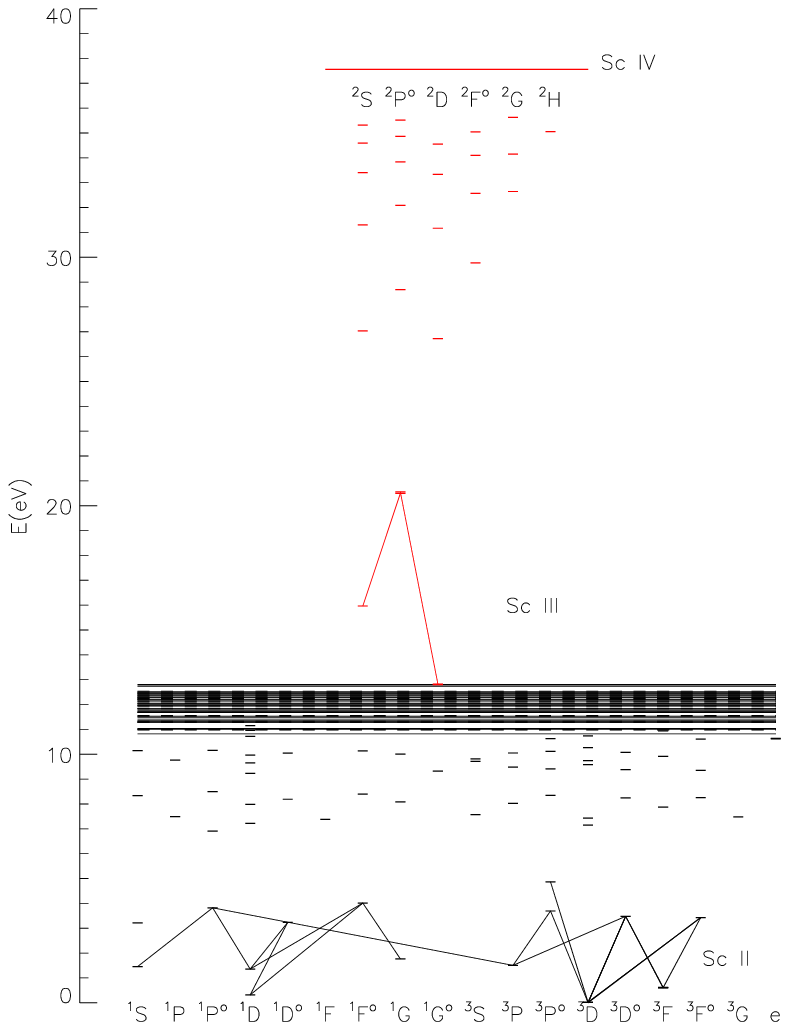}

	\vspace{-9mm}
    \caption{The term structure of Sc\ii\ and Sc\iii, as presented in the model atom. The spectral lines used in Sc abundance analysis arise in the transitions shown as continuous lines.
 The continuous and dashed horizontal lines indicate the energies of the Sc\ii\ even and odd superlevels, respectively. Levels with violation of the LS-coupling are presented in column ’e’.  }\label{fig:atom}
\end{figure*}

\subsection{Energy levels}

The level energies were taken from the National Institute of Standards and Technology (NIST) database\footnote{\url{https://www.nist.gov/pml/atomic-spectra-database}} \citep{NIST_ASD} and the atomic structure calculations by R.~Kurucz\footnote{\url{http://kurucz.harvard.edu/atoms/}} (files c2101e.log and c2101o.log from April 11 of 2011, c2102e.log and c2102o.log from February 16 of 2013).
With the ionization energy $\chi$ = 6.54~eV, scandium is strongly ionized in the atmospheres of A-B type stars. For example, in the model atmosphere with $\Teff$/$\logg$ = 7250/4.20, a fraction of neutral scandium does not exceed 0.3\%. Here, $\logg$ is the surface gravity in logarithmic scale. Three lowest terms of Sc\ione\ are included in the model atom for the particle number conservation.

Singly ionized scandium dominates total number density of Sc in the late A-type atmospheres, however, it becomes the minority species in the early A-type atmospheres. Therefore, the model atom has to ensure effective coupling of the Sc\ii\ levels with the ground state of Sc\iii\ through collisions.
For Sc\ii, 888 energy levels were used to construct the model atom. The multiplet fine structure is neglected except the Sc\ii\ \eu{3d4s}{3}{D}{}{} ground state and the low-excitation term \eu{3d^2}{3}{F}{}{}.
High-excitation levels of Sc\ii\ with a small energy separation and of the same parity were combined into the superlevels. The superlevels include mostly the levels predicted in the atomic structure calculations, but not discovered (so far) in lab measurements. The average energy of the combined superlevel was calculated taking into account the statistical weights of individual levels. The highest levels of Sc\ii\ are separated from the ground state of Sc\iii\ by 0.08--0.26~eV, which is lower than the average kinetic energy of electrons at temperatures of up to 20\,000 K.

NIST provides 43 energy levels of Sc\iii, with an excitation energy of \Eexc\ = 22.84~eV for the highest of them. The atomic structure calculations by R.~Kurucz predict 10 more energy levels for \Eexc\ $\le$ 22.84~eV. The multiplet fine structure is neglected except the Sc\iii\ \eu{3d}{2}{D}{}{} ground state and the low-excitation term \eu{4p}{2}{P}{\circ}{}.

The final model atom includes 3, 134, 27 levels of Sc\ione, Sc\ii, Sc\iii, respectively, and the ground state of Sc\iv. The term diagram for Sc\ii\ and Sc\iii\ is shown in Fig.~\ref{fig:atom}.

\subsection{Radiative transitions}

The model atom includes 4032 and 110 allowed bound-bound (b-b) transitions of Sc\ii\ and Sc\iii, respectively. 
Their transition probabilities were taken from laboratory measurements of \citet[][Sc\ii]{2019ApJS..241...21L} and calculations of R.~Kurucz (files gf2101.lines from April 11 of 2011 and gf2102.lines from February 16 of 2013).
For 21 transitions, which are associated with the low-excitation levels and in which the upper levels can be pumped by ultraviolet radiation, the radiative rates are calculated using the Voigt function for the absorption profile. For other transitions, the Doppler absorption profile is adopted.

For both Sc\ii\ and Sc\iii, the photoionization cross-sections are calculated in the hydrogenic approximation using the effective principal quantum number instead of the principal quantum number. In Sect.~\ref{sect:uncertain}, we check an influence of variations in photoionization cross-sections for Sc\ii\ on the final NLTE results.

\subsection{Collisional transitions}
 
 For 948 transitions between the Sc\ii\ levels with \Eexc $\le$ 9.5~eV, the electron-impact excitations are treated using the effective collision strengths $\Upsilon$ calculated by \citet{2012MNRAS.424.2461G} with the R-matrix method. We note that all the observed lines of Sc\ii\ are formed between the levels in this energy interval (see Fig.~\ref{fig:atom}). For the remaining b-b transitions in Sc\ii\ and Sc\iii, we used the formula of \citet{Reg1962} for allowed transitions and assumed $\Upsilon$ = 1 for  forbidden transitions. 
 
 Electron-impact ionization rate is calculated from the \citet{1962amp..conf..375S} formula with an adopted hydrogenic photoionization cross-section at the threshold.
 
\begin{table*}
	\centering
	\caption{Atmospheric parameters of the sample stars and characteristics of the observed spectra.}
	\label{tab:stars_param}
	\begin{tabular}{llcccrcccllr} %
		\hline\hline \noalign{\smallskip}
\multicolumn{1}{c}{HD,} & Star & $\Teff$ & $\logg$ & Ref & [Fe/H]$^1$ & $\xi_t$ & $V\sin i$ & Ref &\multicolumn{3}{c}{Observed spectra} \\
\cline{10-12}
                             & & [K]  &      &      &  & [\kms] &  [\kms] & & Instrument & Source & $\lambda / \Delta \lambda$ \\
\noalign{\smallskip}\hline \noalign{\smallskip}
 \ 32115    &         & 7250\scriptsize{(100)} & 4.20\scriptsize{(0.05)} & B02  & 0.09\scriptsize{(0.11)} & 2.3\scriptsize{(0.3)} & 9  & B02 & ESPaDOnS      & CAO & 60\,000 \\
 \ 73666    & 40 Cnc  & 9380\scriptsize{(200)} & 3.78\scriptsize{(0.2)} & F07  & 0.24\scriptsize{(0.10)} & 1.8\scriptsize{(0.2)} & 10 & F07 & ESPaDOnS  & F07 & 65\,000 \\
172167      & Vega    & 9550 & 3.95 & CK93 & $-$0.41\scriptsize{(0.17)} & 1.8\scriptsize{(0.5)} & 23.5 & R14 & FOCES   & A. Korn & 40\,000 \\
            &         &      &      &      &         &     &      &     & HIDES   &  T07    & 100\,000 \\
 \ 72660    &         & 9700 & 4.10 & S16  & 0.67\scriptsize{(0.16)} & 1.8 & 7  & R14 & ESPaDOnS  & V. Khalack & 65\,000 \\
 \ 48915    & Sirius  & 9850\scriptsize{(200)} & 4.30\scriptsize{(0.2)} & HL93 & 0.52\scriptsize{(0.06)} & 1.8\scriptsize{(0.5)} & 16 & HL93 & ESPaDOnS & CAO & 65\,000 \\
209459      & 21 Peg  &10400\scriptsize{(200)} & 3.55\scriptsize{(0.1)} & F09  & 0.05\scriptsize{(0.07)} & 0.5\scriptsize{(0.4)} & 3.8 & F09 & ESPaDOnS & CAO & 120\,000 \\
 \ 17081  & $\pi$ Cet &12800\scriptsize{(200)} & 3.75\scriptsize{(0.1)} & F09  & $-$0.08\scriptsize{(0.10)} & 1.0\scriptsize{(0.5)} & 20 & F09 & ESPaDOnS  & F09 & 65\,000 \\
160762  & $\iota$~Her &17500\scriptsize{(200)} & 3.80\scriptsize{(0.05)} & NP12 & 0.00\scriptsize{(0.06)} & 1.0\scriptsize{(0.5)} & 6 & NP12 & ESPaDOnS  & CAO & 65\,000 \\
            &         &      &      &      &         &     &   &      &  STIS     & T. Ayres    & 25\,000 \\
\noalign{\smallskip}\hline \noalign{\smallskip}
\end{tabular}

{\bf Notes.} The numbers in parentheses are the uncertainties in atmospheric parameters. $^1$ NLTE abundances from \citet{2020MNRAS.499.3706M}.
{\bf Ref, Source:} B02 = \citet{2002A&A...389..537B}, F07, F09 = \citet{2007AA...476..911F,2009AA...503..945F}, CK93 = \citet{1993ASPC...44..496C}, S16 = \citet{sitnova_ti},
HL93 = \citet{1993AA...276..142H}, NP12 = \citet{2012A&A...539A.143N}, R14 = \citet{2014A&A...562A..84R}, 
CAO = Common Archive Observation database,
T07 = \citet{2007PASJ...59..245T}.
\end{table*}

\section{Stellar sample, observations and atmospheric parameters}\label{sect:obs}

The developed model atom was tested by analysing the scandium lines in a sample of A-B type stars with well-determined atmospheric parameters, slow rotational velocities ($V\sin i \precsim $ 24~\kms), and high-quality observed spectra available.
The stars were selected from our earlier NLTE studies of chemical species in A–B type stars \citep[the studies are summarised by][]{2020MNRAS.499.3706M}.
Table~\ref{tab:stars_param} lists the stars together with their atmospheric parameters and the sources of observational data.
For consistency with our earlier papers, here, we used for each star exactly the same effective temperature, surface gravity, metallicity ([Fe/H]), microturbulent velocity ($\xi_t$), and the model atmosphere.
The methods of atmospheric parameter determinations were described in detail by \citet{2016MNRAS.462.1123A,2018ApJ...866..153A} and \citet{sitnova_ti} and also in the original papers cited in Table~\ref{tab:stars_param}. We used classical plane-parallel \textsc{LLmodels} model atmospheres \citep{2004AA...428..993S}. For Sirius, the model atmosphere was taken from the website of R.~Kurucz\footnote{\url{http://kurucz.harvard.edu/stars/sirius/ap04t9850g43k0he05y.dat}}.

\begin{figure*}
 \begin{minipage}{170mm}

\hspace{-10mm}
\parbox{0.33\linewidth}{\includegraphics[scale=0.4]{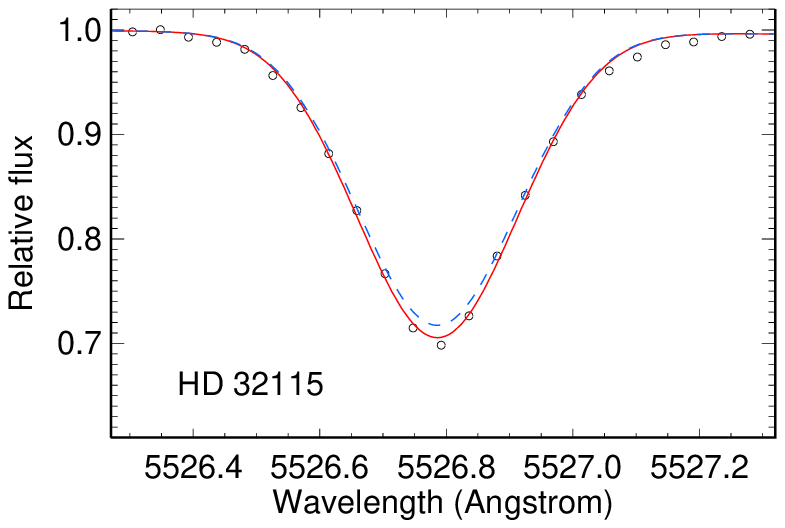}\\
\centering}
\hspace{5mm}
\parbox{0.33\linewidth}{\includegraphics[scale=0.4]{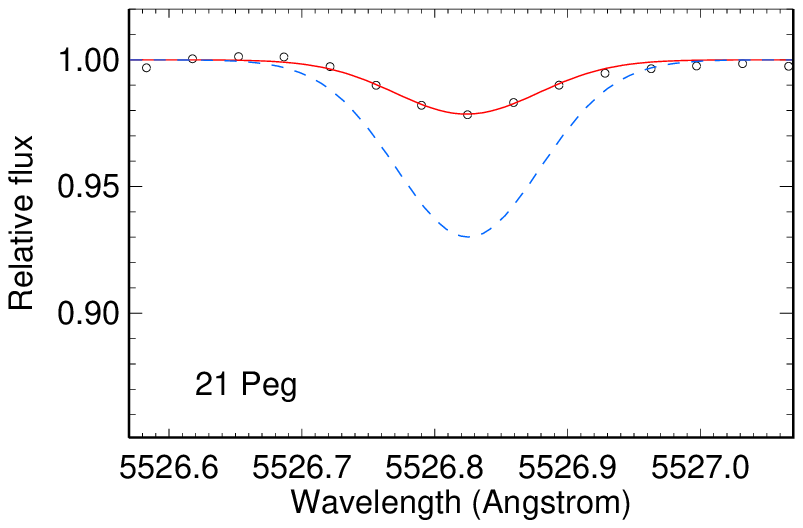}\\
\centering}
\hspace{5mm}
\parbox{0.33\linewidth}{\includegraphics[scale=0.4]{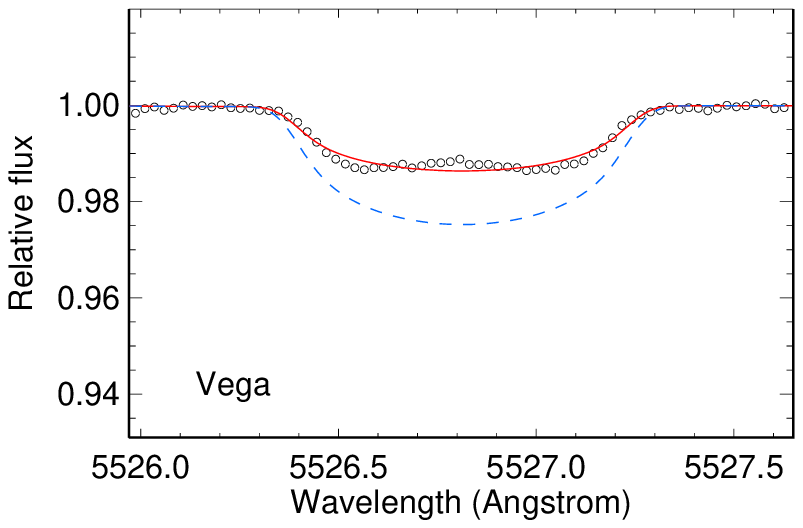}\\
\centering}

	\vspace{-5mm}
    \caption{The best NLTE fits (continuous curve) of Sc\ii\ 5526\,\AA\ in the observed spectra of HD~32115 (left panel), 21~Peg (middle panel), and Vega (right panel). The dashed curves show the LTE profiles computed with the NLTE abundances derived from this line, as indicated in Tables~\ref{tab:hd32115} and \ref{tab:ind_lines}. }
    \label{fig:sc5526}
\end{minipage}
\end{figure*}

We remind briefly that HD~17081, HD~32115, and HD~72660 are primary components of single line spectroscopic binaries (SB1) with negligible flux coming from the secondary star. As in our earlier studies, we assume that it is safe to analyse their spectra ignoring the presence of the secondary components.
We also assume that the atmosphere of a slowly pulsating B-type (the $\beta$~Cephei type) star $\iota$~Her can be represented by a classical hydrostatic model atmosphere. Vega is known as a rapidly rotating star seen nearly pole-on, with an equatorial rotation velocity of $V_e$ = 195~\kms\ and an inclination angle of $i$ = 6.4$^\circ$ according to \citet[][see also Table~1 in their paper]{2021MNRAS.505.1905T}. Temperature and gas pressure on the surface of fast rotator are distributed non-uniformly. From the LTE analysis of five independent series of spectroscopic data and continuum flux, \citet{2010ApJ...712..250H} found the variations in $\Teff$ and $\logg$ over the photosphere, total of 1410~K and 0.26~dex, respectively. In this study, the Vega's atmosphere is represented by a classical homogeneous model atmosphere. We consider that the evaluated NLTE effects can be useful for future advanced studies of the Sc\ii\ lines in Vega. Figure~\ref{fig:sc5526} illustrates a quality of the best fit for Sc\ii\ 5526\,\AA\ in the spectrum of high spectral resolving power ($R = \lambda/\Delta \lambda \simeq$ 100\,000).


The stars HD~32115 (A9~V), HD~73666 (40~Cnc, A1~V), HD~209459 (21~Peg, B9.5~V), HD~17081 ($\pi$~Cet), and HD~160762 ($\iota$~Her, B3~IV) are referred in the literature to as superficially normal ones \citep[][respectively]{2002A&A...389..537B,2007AA...476..911F,2009AA...503..945F,2012A&A...539A.143N}, despite HD~73666 is a Blue Straggler in the Praesepe cluster. \citet{2020MNRAS.499.3706M} show that the NLTE abundances of He, C, O, Na, Mg, Si, Ca, Ti, and Fe in  these stars (when subtracting the metallicity of the Praesepe open cluster for HD~73666) are consistent with the solar abundances, mostly within 0.1~dex, and,
thus, support classification of these stars as superficially normal ones.
Though abundances of the heavy elements Sr, Zr, Ba, and Nd are supersolar, with the highest enhancement for Ba, of up to [Ba/H] = 0.98 in 21~Peg \citep{2020MNRAS.499.3706M}.

HD~48915 (Sirius) and HD~72660 reveal supersolar abundances of the Fe group elements \citep[for example,][respectively]{1989PASJ...41..279S,1999A&A...341..233V}, and they are referred in the literature to as Am stars.
Based on the NLTE abundance analysis, \citet{2020MNRAS.499.3706M} have confirmed that these two Am stars are Fe rich ([Fe/H] $\simeq$ 0.5) and reveal strong enhancements of up to 1.5~dex in the heavy elements Sr, Zr, Ba, and Nd.

HD~172167 (Vega) was identified by \citet{1990ApJ...363..234V} as a mild $\lambda$~Bootes type star. \citet{2020MNRAS.499.3706M} have shown that Vega has close-to-solar NLTE abundances of C and O and subsolar NLTE abundances of Mg to Fe, at the level of [Element/H] $\simeq$ 0.4~dex.

Characteristics of the observed spectra are listed in Table~\ref{tab:stars_param}. For all the stars except Vega, we used the spectra observed with the ESPaDOnS instrument\footnote{\url{http://www.cfht.hawaii.edu/Instruments/Spectroscopy/Espadons/}} of the Canada-France-Hawaii Telescope (CHFT). They have $R >$ 60\,000 and a signal-to-noise ratio of S/N $>$ 200.
For Vega, we used two spectra. One of them was observed by A. Korn with the spectrograph FOCES (fibre optics Cassegrain echelle spectrograph) at the 2.2-m telescope of the Calar Alto Observatory. Another one was obtained by \citet{2007PASJ...59..245T} with the HIgh-Dispersion Echelle Spectrograph (HIDES) at the coud{\'{e}} focus of the 188~cm reflector at Okayama Astrophysical Observatory\footnote{\url{http://pasj.asj.or.jp/v59/n1/590122/590122-frame.html}}.

Abundance analysis of $\iota$~Her takes an advantage of using its UV spectrum in the 1160-3046\,\AA\ range obtained on the Hubble Space Telescope with the STIS spectrograph and
 provided by Thomas Ayres on the ASTRAL project website\footnote{\url{http://casa.colorado.edu/~ayres/ASTRAL/}}.

\begin{figure}
	\includegraphics[width=\columnwidth]{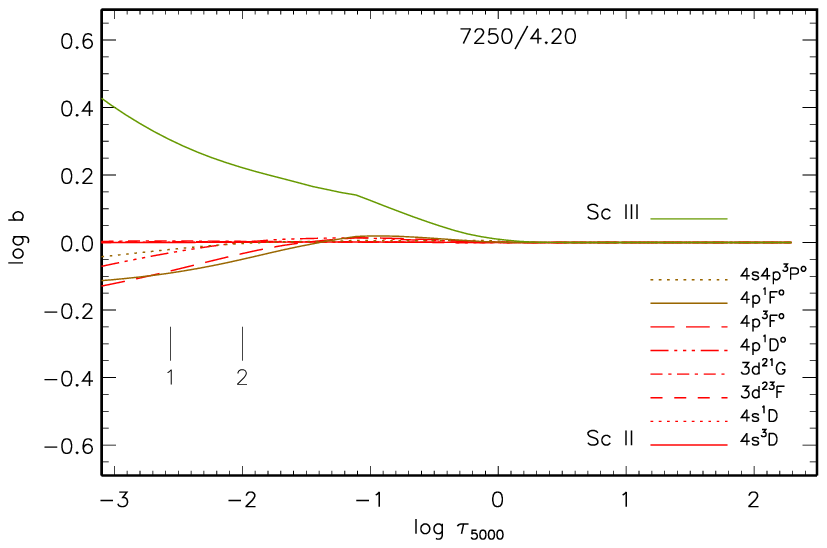}
	
	\vspace{-5mm}
	\includegraphics[width=\columnwidth]{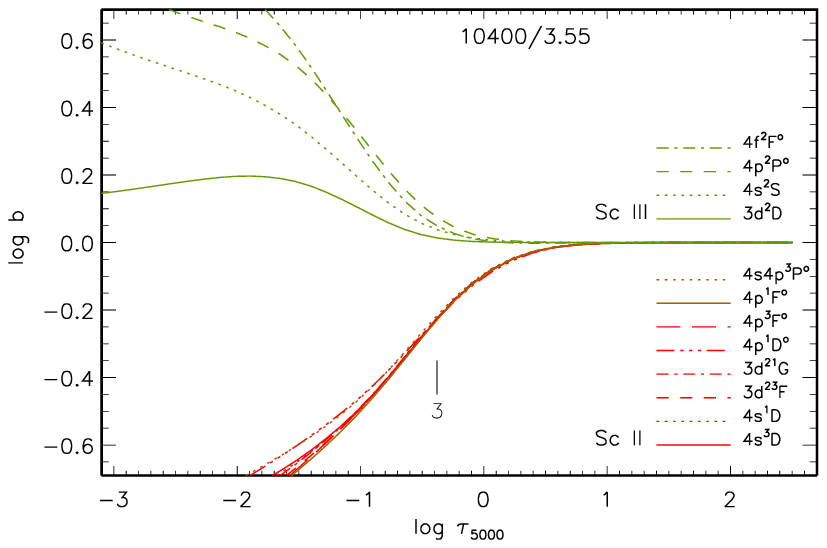}

	\vspace{-5mm}
	\includegraphics[width=\columnwidth]{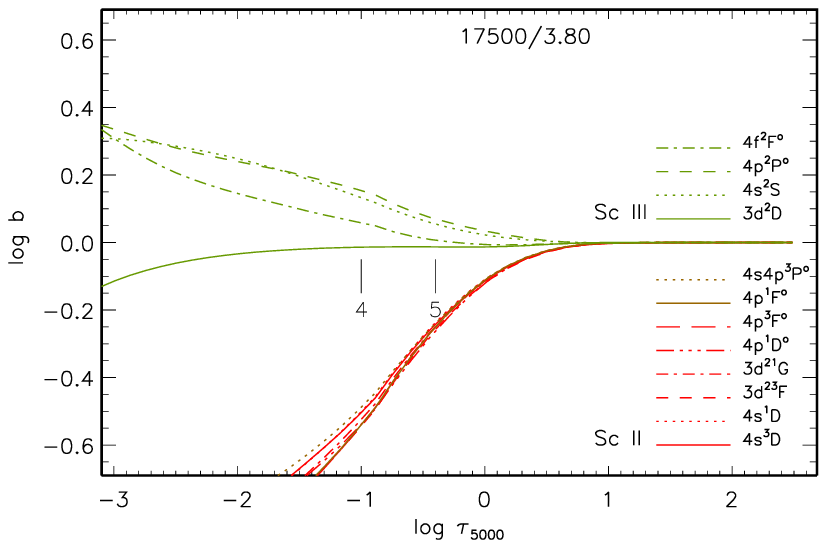}

	\vspace{-5mm}
    \caption{Departure coefficients, log~b, for the levels of Sc\ii\ (red and brown curves) and Sc\iii\ (green curves) as a function of $\log \tau_{5000}$ in the model atmospheres 7250/4.20/0, 10400/3.55/0, and 17500/3.80/0.02. The selected levels are quoted in the right part of each panel. Tick marks indicate the locations of line center optical depth unity for the following lines: Sc\ii\ 4400 (1), 5526 (2), 4246\,\AA\ (3) and Sc\iii\ 1603 (4), 2734\,\AA\ (5).}
    \label{fig:bfactors}
\end{figure}

\section{Statistical equilibrium of scandium and NLTE effects on spectral lines}\label{sect:nlte}

We used a modified version of the {\sc DETAIL} code \citep{Giddings81,Butler84,2011JPhCS.328a2015P} to solve the coupled radiative transfer and statistical equilibrium (SE) equations.
Figure~\ref{fig:bfactors} displays the departure coefficients, ${\rm b = n_{NLTE}/n_{LTE}}$, in the model atmospheres with different $\Teff$.  Here,
${\rm n_{NLTE}}$ and ${\rm n_{LTE}}$ are the statistical equilibrium and thermal (Saha-Boltzmann) number densities, respectively.

Ionization state of scandium changes dramatically when moving along a sequence of the models representing the atmospheres of our sample stars. Sc\ii\ is a majority species in the coolest model, $\Teff$/$\logg$ = 7250/4.20. In the line-formation layers, its number density, $N$(Sc\ii), exceeds that for Sc\ione\ and Sc\iii, by more than 2~dex.
As a result, the ground state of Sc\ii\ keeps its thermodynamic equilibrium (TE) population ($\log$~b = 0) throughout the atmosphere and the low-excitation levels do the same up to the reference optical depth $\log \tau_{5000} \simeq -1.5$ (Fig.~\ref{fig:bfactors}, top panel). The Sc\ii\ lines observed in HD~32115 are strong enough, and their cores form outward $\log \tau_{5000} \simeq -2$, where the upper levels of the corresponding transitions are slightly underabundant relative to their TE populations ($\log$~b $< 0$) resulting in dropping the line source function relative to the Planck function and strengthened lines. The effect is small, as illustrated by Fig.~\ref{fig:sc5526} (left panel) for Sc\ii\ 5526~\AA\ (transition \eu{3d^2}{1}{G}{}{} -- \eu{4p}{1}{F}{\circ}{}).

With increasing $\Teff$, Sc\ii\ is ionized and the Sc\iii\ number density increases. Fractions of Sc\ii\ and Sc\iii\ are approximately equal in the 9550/3.95 model, and $N$(Sc\iii) exceeds $N$(Sc\ii) in the 10400/3.55 model, by more than one order of magnitude.
Superthermal UV radiation of a non-local origin below the thresholds of the Sc\ii\ levels with \Eexc\ of 3.2 to 4.8~eV ($\lambda_{\rm thr}$ = 1294 to 1562\,\AA) leads to an overionization of Sc\ii\ (log~b $< 0$, Fig.~\ref{fig:bfactors}, middle panel) resulting in weakened lines of Sc\ii\ (Fig.~\ref{fig:sc5526}, middle panel for Sc\ii\ 5526~\AA).

\begin{figure}
	\includegraphics[width=\columnwidth]{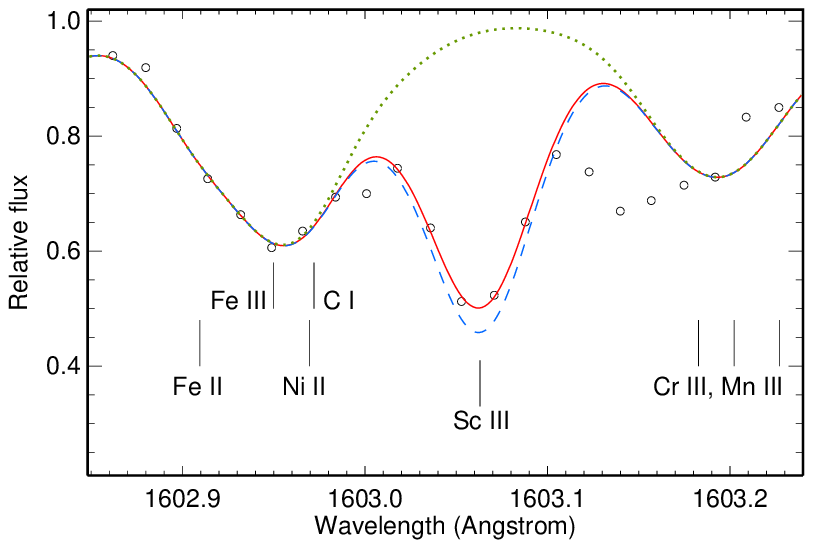}

	\vspace{-5mm}
	\includegraphics[width=\columnwidth]{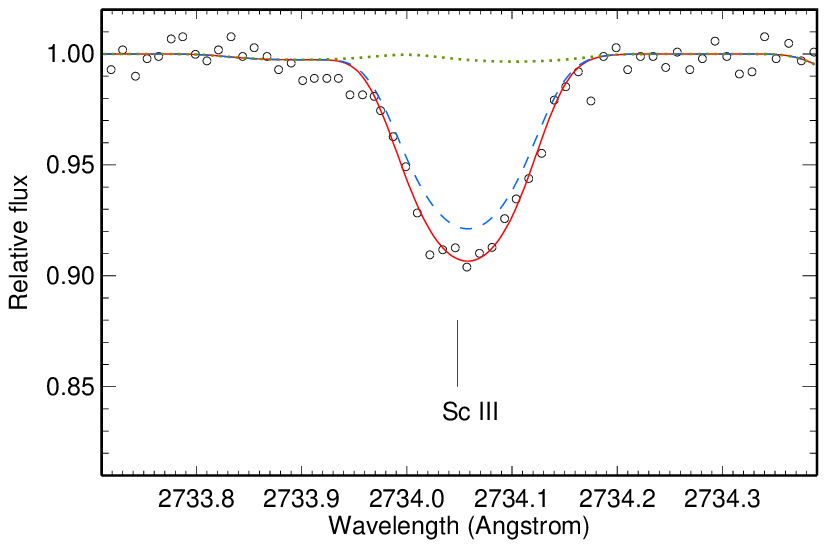}

	\vspace{-5mm}
    \caption{The best NLTE fits (continuous curve) of Sc\iii\ 1603 (top panel) and 2734\,\AA\ (bottom panel) in the observed spectrum of $\iota$~Her. The dashed curves show the LTE profiles computed with the NLTE abundances derived from the corresponding lines and indicated in Table~\ref{tab:hd32115}. The dotted curves show the synthetic spectrum with no scandium in the atmosphere. }
    \label{fig:sc3lines}
\end{figure}

Sc\ii\ is a minority species in the hottest model, 17500/3.80, with a fraction of less than 0.001, and is subject to a strong overionization (Fig.~\ref{fig:bfactors}, bottom panel). The majority species is Sc\iii, and its ground state has close to the TE population. Strong pumping transition \eu{3d}{2}{D}{}{}  -- \eu{4p}{2}{P}{\circ}{} (1598, 1603, 1610~\AA) produces an overpopulation (log~b $> 0$) of the Sc\iii\ \eu{4p}{2}{P}{\circ}{} level and the other excited levels via the transitions from \eu{4p}{2}{P}{\circ}{}. As a result, the Sc\iii\ resonance line at 1603~\AA\ is weakened compared to the LTE case, while the 2734~\AA\ line arising in the Sc\iii\ \eu{4s}{2}{S}{}{}  -- \eu{4p}{2}{P}{\circ}{} transition is strengthened (Fig.~\ref{fig:sc3lines}). In the next section, we show that NLTE removes an abundance difference of 0.29~dex obtained in LTE between the Sc\iii\ subordinate and resonance lines.

\section{Abundance results}\label{sect:abund}

\subsection{Line list and analysis of stellar spectra}

\begin{table}
\centering
\caption{NLTE and LTE abundances, $\eps{}$, from individual lines of scandium in HD~32115, Sirius, $\pi$~Cet, and $\iota$~Her.}
\label{tab:hd32115}
\begin{tabular}{lcrcc} %
	\hline\hline \noalign{\smallskip}
\multicolumn{1}{c}{ $\lambda$} & \Eexc & log $gf$ & LTE & NLTE \\ 
\multicolumn{1}{c}{ [\AA]}     & [eV]  &          &     &  \\
\noalign{\smallskip}\hline \noalign{\smallskip}
\multicolumn{5}{l}{ \ \ HD~32115} \\
Sc\ii\ 4294.77 & 0.61 & --1.36 &   2.87 & 2.86 \\
Sc\ii\ 4400.39 & 0.61 & --0.54 &  3.19 &  3.13  \\
Sc\ii\ 4415.56 & 0.60 & --0.68 &  3.25 &  3.20  \\
Sc\ii\ 5031.02 & 1.36 & --0.41 &  3.11 &  3.11  \\
Sc\ii\ 5239.81 & 1.46 & --0.76 &  3.03 &  3.04  \\
Sc\ii\ 5526.79 & 1.77 & --0.01 &  3.22 &  3.16  \\
Sc\ii\ 5641.00 & 1.50 & --0.99 &  3.05 &  3.05 \\
Sc\ii\ 6245.64$^1$ & 1.51 & --1.02 & 2.91 & 2.89 \\
\multicolumn{1}{r}{Mean $\eps{}$}  &  &      &  3.10(0.14) &  3.05(0.12) \\
\multicolumn{1}{r}{[Sc/H]}    &      &      & 0.06        &  0.01 \\
 \multicolumn{5}{l}{ \ \ Sirius} \\
Sc\ii\ 4246.82  &  0.31 & 0.24 & 1.88(0.03) & 2.24(0.03) \\
\multicolumn{1}{r}{[Sc/H]} &      &        & --1.16     &  --0.80 \\
 \multicolumn{5}{l}{ \ \ $\pi$ Cet} \\
Sc\ii\ 4246.82  &  0.31 & 0.24 & 2.67(0.1) & 3.19(0.1) \\
\multicolumn{1}{r}{[Sc/H]}  &      &        & --0.37    &  0.15 \\
 \multicolumn{5}{l}{ \ \ $\iota$~ Her} \\
Sc\iii\ 1603.06 & 0.02 & --0.28  & 2.89 & 3.05  \\
Sc\iii\ 1610.19 & 0.00 & --0.54  & 2.84 & 2.98  \\
Sc\iii\ 2699.07 & 3.17 &   0.08  & 3.13 & 3.00  \\
Sc\iii\ 2734.05 & 3.17 & --0.23  & 3.17 & 3.06  \\
\multicolumn{1}{r}{Mean $\eps{}$} &   &    &  3.03(0.17) &  3.02(0.04) \\
\multicolumn{1}{r}{[Sc/H]} &      &        & --0.01    &  --0.02 \\
\noalign{\smallskip}\hline \noalign{\smallskip}
	\end{tabular}
	
{\bf Notes.} $^1$ $gf$ from R.~Kurucz calculations. The numbers in parentheses are the statistical errors, see  text for their description.
\end{table}

\begin{table*}
\centering
\caption{NLTE and LTE abundances, $\eps{}$, from individual lines of Sc\ii\ in the sample stars.}
\label{tab:ind_lines}
	\begin{tabular}{ccrccccccccccc} %
		\hline\hline \noalign{\smallskip}
\multicolumn{1}{c}{ $\lambda$} & \Eexc & log $gf$ & \multicolumn{2}{c}{HD~73666} & & \multicolumn{2}{c}{Vega} & & \multicolumn{2}{c}{HD~72660} & &  \multicolumn{2}{c}{21 Peg} \\
\multicolumn{1}{c}{ [\AA]}     & [eV]  &          & LTE  & NLTE & & LTE  & NLTE & & LTE  & NLTE & &  LTE  & NLTE \\
\noalign{\smallskip}\hline \noalign{\smallskip}
 4246.82 & 0.31 &  0.24  & 3.02 & 3.21 & & 2.43 & 2.80 & & 2.56 & 2.81 & &  2.53 &   3.15 \\
 4374.46 & 0.62 & --0.46 & 3.05 & 3.26 & &      &      & & 2.70 & 2.91 & &  2.68 &   3.26 \\
 4400.39 & 0.61 & --0.54 & 3.03 & 3.24 & & 2.50 & 2.84 & & 2.65 & 2.86 & &  2.57 &   3.17 \\
 4415.56 & 0.60 & --0.68 & 3.03 & 3.25 & &      &      & & 2.64 & 2.86 & &  2.69 &   3.27 \\
 5031.02 & 1.36 & --0.41 & 3.03 & 3.24 & & 2.50 & 2.83 & & 2.62 & 2.82 & &  2.57 &   3.15 \\
 5526.79 & 1.77 & --0.01 & 3.09 & 3.27 & & 2.64 & 2.94 & & 2.69 & 2.88 & &  2.67 &   3.19 \\
\multicolumn{2}{l}{Mean $\eps{}$} &        & 3.04 & 3.25 & & 2.52 & 2.85 & & 2.65 & 2.86 & &  2.62 &   3.20 \\
 ($\sigma$) &   &        & (0.03)& (0.02)& & (0.09)& (0.06) & &(0.05)&(0.04) & & (0.07)&(0.05) \\
 \ [Sc/H] &     &        & 0.00 & 0.21 & & --0.52 & --0.19 & & --0.39 & --0.18 & & --0.42 & 0.16 \\
\noalign{\smallskip}\hline \noalign{\smallskip}
\end{tabular}
\end{table*}

For seven stars, the Sc abundances were derived from the Sc\ii\ lines, which were selected from the line list employed in our earlier study (MR2022). Their oscillator strengths were taken from the laboratory measurements of \citet{2019ApJS..241...21L}. The exception is Sc\ii\ 6245~\AA, for which we adopted $gf$-value from calculations of R.~Kurucz.
No Sc\ii\ lines were detected in spectrum of $\iota$~Her, however, we could measure the four relatively unblended lines of Sc\iii\ in the UV spectral range. Their $gf$-values were calculated by \citet{Weiss_sc3_1967} and R.~Kurucz (file gf2102.lines, as cited above). We prefer to apply the more recent data of R.~Kurucz.
The lines and their atomic parameters are listed in Tables~\ref{tab:hd32115} and \ref{tab:ind_lines}. The van der Waals broadening and quadratic Stark effect broadening constants were predicted by R.~Kurucz and available in the VALD (Vienna Atomic Line
Database) database \citep{2015PhyS...90e4005R}.

Scandium is represented in the nature by the only stable isotope, $^{45}$Sc. Nucleon-electron spin interactions in this isotope lead to hyper-fine splitting (HFS) of the energy
levels, resulting in absorption lines divided into multiple components. For lines of Sc\ii, wavelengths and oscillator strengths of the HFS components were taken from the updated VALD database \citep{2019ARep...63.1010P}.
No HFS data are available for lines of Sc\iii.


Analyses of observed spectra were based on line profile fitting. The synthetic spectra were calculated with the {\sc Synth}V\_NLTE code \citep{2019ASPC}, which implements the pre-computed departure coefficients from the {\sc DETAIL} code for the chemical species under investigation and treats spectral lines of the other chemical species under the LTE assumption.
The line list and atomic data for the synthetic spectra computations were taken from the VALD database.
The best fit to the observed spectrum was achieved automatically using the {\sc IDL binmag} code \citep{2018ascl.soft05015K}.

Tables~\ref{tab:hd32115} and \ref{tab:ind_lines} present the LTE and NLTE abundances from individual lines and the mean Sc abundances. Hereafter, the [Sc/H] values are calculated using the solar system Sc abundance $\eps{met}$ = 3.04$\pm$0.03 from \citet{2021SSRv..217...44L}.

We obtained that the difference between the NLTE and LTE abundances referred to as the NLTE abundance correction, $\Delta_{\rm NLTE} = \eps{NLTE} - \eps{LTE}$, is minor or slightly negative, down to $\Delta_{\rm NLTE} = -0.06$~dex, for different lines of Sc\ii\ in HD~32115. The NLTE corrections are positive for the hotter stars and grow with increasing $\Teff$, up to $\Delta_{\rm NLTE}$ = 0.62~dex for Sc\ii\ 4246~\AA\ in 21~Peg.

\subsection{Uncertainties in derived abundances}\label{sect:uncertain}

If more than one scandium line were measured ($N_l > 1$), the statistical error of the mean abundance was calculated as the dispersion in the single line measurements around the mean, $\sigma = \sqrt{\Sigma(\overline{x}-x_i)^2 / (N_l-1)}$. For Sirius and $\pi$~Cet, with the only measured line, Sc\ii\ 4246~\AA, we adopted an uncertainty of 0.1\%\ in the continuum placement as the source of the abundance error, and the statistical errors were estimated as 0.03~dex and 0.10~dex, respectively.

For each star with $N_l > 1$, NLTE leads to smaller $\sigma$. We note, in particular, $\iota$~Her with $\sigma$ = 0.17~dex and 0.04~dex
in LTE and NLTE, respectively. This is owing to NLTE
reducing absorption in the resonance lines, but increasing that for the subordinate lines compared to the LTE case.

\begin{table}
\centering
\caption{Error estimates for NLTE calculations of the Sc\ii\ lines in 21~Peg.}
	\label{tab:uncertainties}
	\begin{tabular}{llccc} %
	\hline\hline \noalign{\smallskip}
 & & \multicolumn{3}{c}{Changes in $\eps{}$ (dex)} \\
\cline{3-5}
 & & 4246\,\AA & 4400\,\AA & 5526\,\AA \\
\noalign{\smallskip}\hline \noalign{\smallskip}
 \multicolumn{5}{c}{NLTE treatment} \\
 NLTE -- LTE & $\Delta_{\rm NLTE}$ & 0.62 & 0.60 & 0.52  \\
 \multicolumn{5}{l}{Photoionizations:} \\
 cross-sections \texttimes 0.1 & $\sigma_{\rm RBF}$ & $-0.06$ & $-0.05$ & $-0.05$ \\
 \multicolumn{5}{l}{Opacity below 1507~\AA :} \\
 Si abundance \texttimes 3.16 & $\sigma_{\rm UVop}$ & $-0.07$ & $-0.04$ & $-0.06$ \\
 \multicolumn{5}{l}{Electron-impact ionizatios} \\
 cross-sections \texttimes 10 & $\sigma_{\rm CBF}$ & $-0.01$ & $ 0.00$ & $0.00$ \\
 \multicolumn{5}{l}{Electron-impact excitations} \\
 $\Upsilon$ \texttimes 0.5 & $\sigma_{\rm CBB}$ & +0.01 & +0.01 & $0.00$ \\
 $\Upsilon$ \texttimes 2   &                    & $-0.02$ & $-0.02$ & $0.00$ \\
 \multicolumn{5}{c}{Atmospheric parameters} \\
 $\Teff$ --200~K & $\sigma_{\Teff}$ & $-0.15$ & $-0.16$ & $-0.14$  \\
 $\logg$ +0.1    & $\sigma_{\logg}$ & $-0.02$ & $-0.02$ & $-0.02$  \\
 $\xi_t$ +0.4~\kms & $\sigma_{\xi}$ & $-0.01$ & $ 0.00$ & $ 0.00$ \\
\noalign{\smallskip}\hline \noalign{\smallskip}
\multicolumn{5}{l}{{\bf Notes.} 0.00 means smaller than 0.01~dex, in absolute value. }
\end{tabular}
\end{table}

 Systematic abundance errors caused by the uncertainties in NLTE treatment and atmospheric parameters were evaluated for the selected lines in 21~Peg, which reveal the strongest NLTE effects. Table~\ref{tab:uncertainties} summarises results of our tests.

Since the NLTE effects on Sc\ii\ in the atmosphere of 21~Peg are mainly caused by overionization of the Sc\ii\ levels with \Eexc\ of 3.2 to 4.8~eV (Sect.~\ref{sect:nlte}), the uncertainties in radiative rates of the bound-free (b-f) transitions from these levels are expected to be the main source of the uncertainties in NLTE treatment. For any b-f transition, its radiative rate depends on the photoionization cross-sections and the intensity of ionizing radiation and, thus, on the backgroung opacity below the ionization threshold. A reduction of the hydrogenic photoionization cross-sections by one order of magnitude leads to the weaker NLTE effects for  Sc\ii\ and the lower NLTE abundances derived, by up to 0.06~dex. Variations in the backgroung opacity below 1507~\AA\ were simulated by variations in the Si abundance. Silicon was selected because photoionization of the Si\ii\ \eu{4s}{2}{S}{}{} ($\lambda_{\rm thr}$ = 1507\,\AA) and \eu{3p^2}{2}{D}{}{} ($\lambda_{\rm thr}$ = 1306\,\AA) levels produces a notable contribution to the backgroung opacity in the 1294 to 1562\,\AA\ spectral range, where the thresholds of the important Sc\ii\ levels are located. Notable reduction of the NLTE abundance of Sc, by up to 0.07~dex, can be obtained, if the Si abundance is increased by 0.5~dex compared to the solar value.
Variations in the collisional rates produce minor abundance shifts of no more than 0.02~dex.

\citet{2009AA...503..945F} estimate the uncertainties in $\Teff$, $\logg$, and $\xi_t$ as 200~K, 0.1~dex, and 0.4~\kms, respectively. A downward revision of $\Teff$ would reduce the mean abundance by 0.15~dex. Upward revision of $\logg$ would act in the same direction, however, the abundance shift is small, $-0.02$~dex. Variations in $\xi_t$ nearly do not affect the derived abundance because the Sc\ii\ lines are weak in 21~Peg.

\begin{figure}
	\includegraphics[width=\columnwidth]{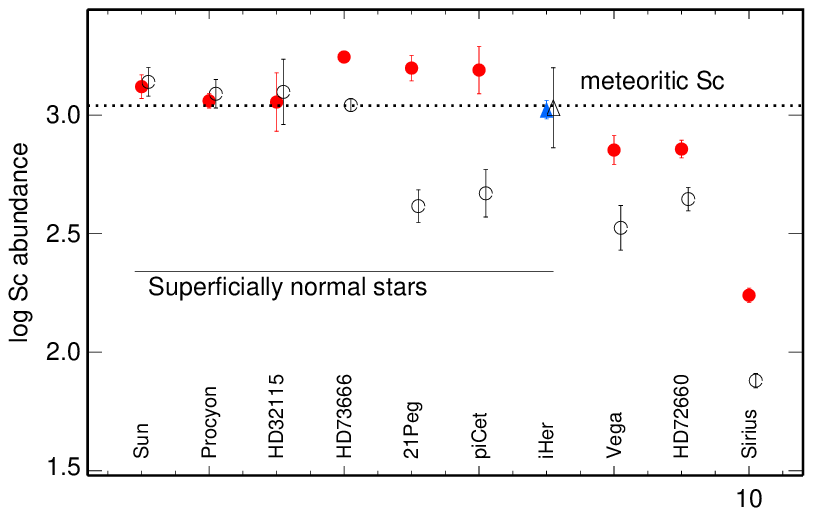}

	\vspace{-5mm}
    \caption{The LTE (open symbols) and NLTE (filled symbols) abundances of scandium, $\eps{}$, in the investigated stars. The circles and triangles correspond to the Sc\ii\ and Sc\iii -based abundances, respectively. The dotted line marks the meteoritic abundance from \citet{2021SSRv..217...44L}.}
    \label{fig:sc_abund}
\end{figure}

\subsection{Scandium in the superficially normal stars and Am stars}

Figure~\ref{fig:sc_abund} displays the mean abundances of the sample stars and, for comparison, of the Sun and a benchmark F-type star HD~61421 (Procyon) that reveals close-to-solar chemical abundances \citep{2020A&A...642A.182A}. The NLTE abundances of Sc in the Sun and Procyon were derived by MR2022 using the model atom of Sc\ii.
It can be seen that abundances of Procyon, HD~32115, and $\iota$~Her are consistent with the solar system Sc abundance. The NLTE abundance of HD~73666 is higher, by 0.21~dex, however, this can be explained by a supersolar metallicity of the Praesepe cluster: [Fe/H] = 0.14 and 0.11 according to \citet{2003AJ....125.1397C} and \citet{2008A&A...483..891F}, respectively. \citet{2020MNRAS.499.3706M} derived [Fe/H] = 0.24 for HD~73666. For 21~Peg and $\pi$~Cet, NLTE removes substantial discrepancies with the solar Sc abundance found in the LTE analysis. Their NLTE abundances, [Sc/H] = 0.16 and 0.15, are a little bit high, however, they agree with both $\eps{met}$ = 3.04$\pm$0.03 and the solar photosphere NLTE abundance, $\eps{\odot}$ = 3.12$\pm$0.05 (MR2022), when
taking into account both statistical and systematic errors discussed in Sect.~\ref{sect:uncertain}.

Different studies report the higher solar photosphere Sc abundance compared to the meteoritic one, by more than 1-2$\sigma$: $\eps{\odot}$ = 3.16$\pm$0.03 \citep[][1D-LTE]{2019ApJS..241...21L}, 3.14$\pm$0.04 \citep[][1D, NLTE]{2021A&A...653A.141A}, 3.17$\pm$0.04 \citep[][3D, NLTE]{2021A&A...653A.141A}, and 3.12$\pm$0.05 (MR2022, 1D-NLTE). Here, 1D and 3D denote calculations with the classical plane-parallel model atmosphere and the hydrodynamic calculations, respectively. Self-consistent NLTE analysis of the Sc\ii\ lines with a 3D model atmosphere is not performed yet. Denotations 1D, NLTE and 3D, NLTE mean that the NLTE abundance corrections calculated by \citet{Zhang2008_sc} with the solar 1D model atmosphere were added to the 1D-LTE and 3D-LTE abundances. \citet{nlte_sc2} raised  questions of (i) what is the source of discrepancies between the solar photosphere and the solar system Sc abundances and (ii) whether the solar photosphere abundance or the meteoritic one should be referred to as the cosmic abundance standard for scandium.

Our results for the superficially normal stars favour the meteoritic Sc abundance as the cosmic abundance standard, however, a cosmic scatter of the Sc abundance, at the level of 0.1~dex, cannot be ruled out. Scandium abundances of an extended sample of the superficially normal stars need to be determined to fix the cosmic abundance standard for scandium.

NLTE reduces a deficiency of Sc in the atmospheres of our two Am stars, HD~72660 and Sirius, compared to the solar Sc abundance, 
however, their Sc abundances remain to be subsolar.


\begin{figure}
	\includegraphics[width=\columnwidth]{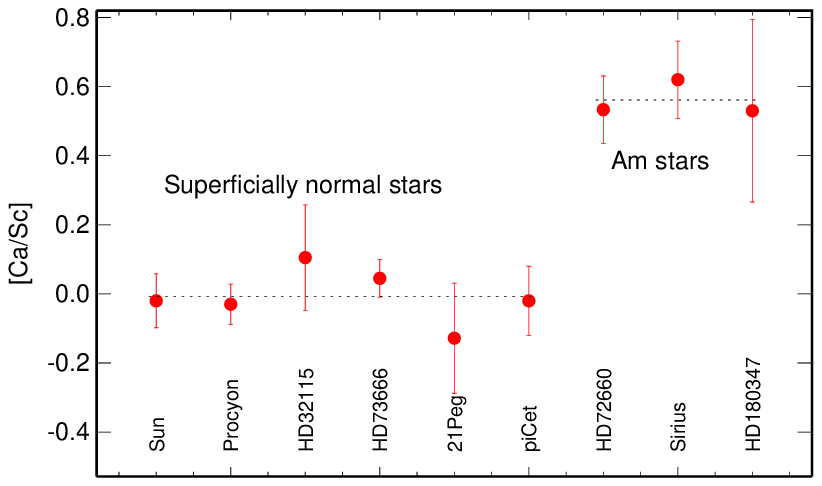}

	\vspace{-5mm}
    \caption{The NLTE abundance ratios [Ca/Sc] of the superficially normal and Am stars. The dotted lines mark the mean values, [Ca/Sc] = $-0.01\pm 0.08$ and 0.56$\pm$0.05, respectively.}
    \label{fig:casc}
\end{figure}

\begin{table}
\centering
\caption{Abundances of Ca, Sc, and Sr in the Am stars.}
	\label{tab:cascsr}
	\begin{tabular}{rccrrcl} %
	\hline\hline \noalign{\smallskip}
 \multicolumn{1}{c}{HD} & $\Teff$ & [Fe/H] & [Ca/H] & [Sc/H] & [Sr/H] & Source\\
 \noalign{\smallskip}\hline \noalign{\smallskip}
 \multicolumn{7}{c}{NLTE calculations} \\
 48915 & 9850 & 0.52 & -0.18 & -0.80 &  0.95 & M20, TS \\
 72660 & 9700 & 0.67 &  0.35 & -0.18 &  1.47 & M20, TS \\
180347 & 7740 & 0.22 & -0.87 & -1.40 &  0.93 & TMJ23     \\
 \multicolumn{7}{c}{Literature LTE calculations} \\
209625 & 7700 & 0.12 & -0.42 & -1.07 &  0.82 & ACK97 \\
22615  & 8410 & 0.15 &  0.37 & -0.39 &  0.44 & GM08  \\
23325  & 7640 & 0.34 & -0.14 & -0.39 &  0.16 & GM08  \\
23631  & 9610 & 0.26 & -0.19 & -0.92 &  0.34 & GM08  \\
27628  & 7310 & 0.07 & -0.36 & -1.09 &  0.55 & GVM10 \\
27962  & 9025 & 0.32 &  0.02 & -0.90 &  0.84 & GVM10 \\
28226  & 7465 & 0.31 &  0.11 & -0.62 &  0.32 & GVM10 \\
28355  & 7965 & 0.35 &  0.00 & -1.05 &  0.78 & GVM10 \\
28546  & 7765 & 0.11 &  0.19 & -0.56 &  0.65 & GVM10 \\
30210  & 8080 & 0.51 & -0.27 & -0.93 &  0.91 & GVM10 \\
 97633 & 9330 & 0.20 &  0.09 & -0.18 &  0.83 & AGH15 \\
214994 & 9535 & 0.37 & -0.14 & -0.16 &  1.09 & AGH15 \\
 \noalign{\smallskip}\hline \noalign{\smallskip}
\end{tabular}

{\bf Notes.} M20 = \citet{2020MNRAS.499.3706M}, TS = This study, TMJ23 = \citet{2023MNRAS.524.1044T}, ACK97 = \citet{1997MNRAS.288..470A}, GM08 = \citet{2008A&A...483..567G}, GVM10 = \citet{2010A&A...523A..71G}, AGH15 = \citet{2015PASP..127...58A}.
\end{table}

\subsection{Ca/Sc abundance ratio is a classification criterium of Am stars}

Abundances of scandium and calcium are often used in classifying a star as Am.
Currently, accurate NLTE abundances of both elements are available for three Am stars. These are Sirius and HD~72660 in our sample and HD~180347 from \citet{2023MNRAS.524.1044T}. Despite large abundance differences between these three stars for both Ca and Sc, for example, [Sc/H] = $-0.80$ (Sirius), $-0.18$ (HD~72660), and $-1.40$ (HD~180347),
they reveal very similar overabundances of Ca relative to Sc, with [Ca/Sc] = 0.62, 0.53, and 0.53, respectively (Fig.~\ref{fig:casc}). The Ca NLTE abundances of our sample stars are taken from \citet{2018MNRAS.477.3343S}. For the Sun and Procyon, we use the NLTE abundances, $\eps{Ca}$ = 6.33$\pm$0.06 and 6.26$\pm$0.05, respectively, from \citet{2017AA...605A..53M}. As expected, in six superficially normal stars shown in Fig.~\ref{fig:casc}, the Ca/Sc ratio is on average solar: [Ca/Sc] = $-0.01\pm 0.08$. The star $\iota$~Her was not included in Fig.~\ref{fig:casc} and in the mean Ca/Sc ratio due to the uncertainty in its Ca abundance \citep[see][for discussion]{2018MNRAS.477.3343S}. 

\begin{figure}
	\includegraphics[width=\columnwidth]{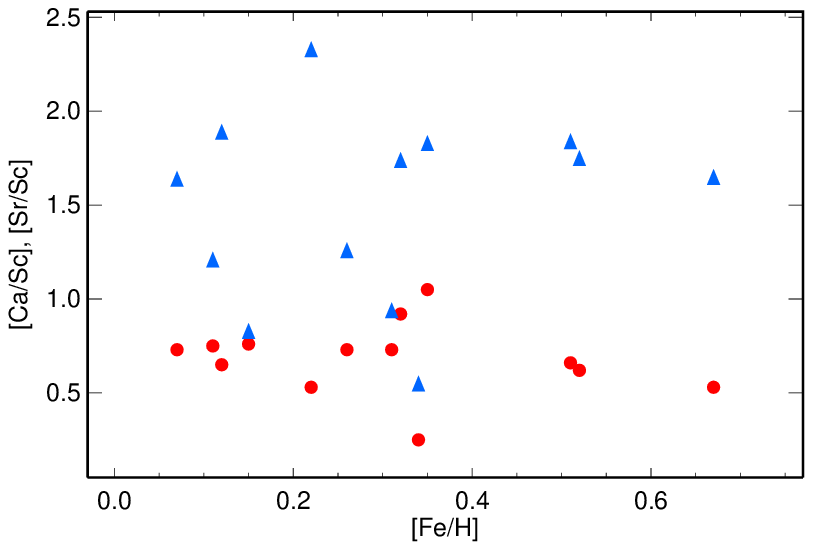}
	\includegraphics[width=\columnwidth]{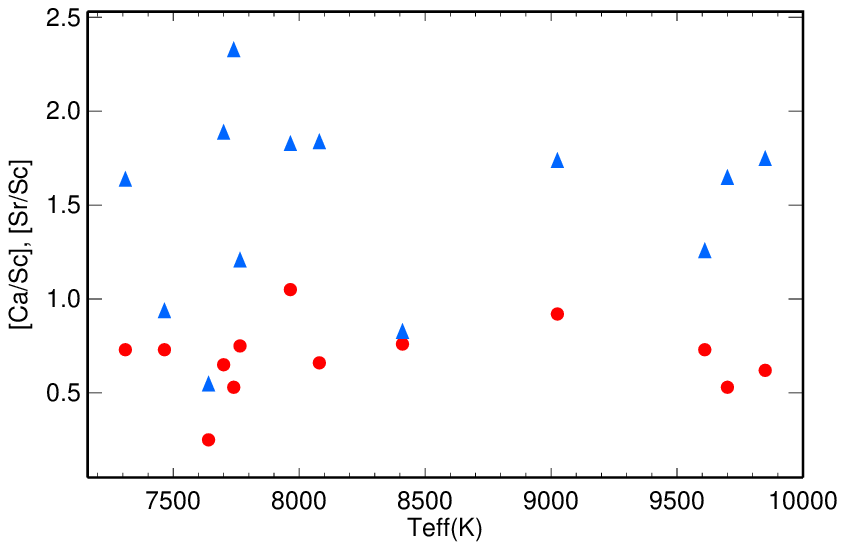}

	\vspace{-5mm}
    \caption{[Ca/Sc] (red circles) and [Sr/Sc] (blue triangles) abundance ratios of the Am stars as functions of metallicity and effective temperature. See Table~\ref{tab:cascsr} for the sources of data. }
    \label{fig:casc_lit}
\end{figure}

In order to justify our result for [Ca/Sc] in Am stars, we increased the statistics of Am stars by compiling the elemental abundances from the literature. Since they all were determined under the LTE assumption, we selected the Am stars with $\Teff <$ 9000~K. In the next section, we show that, in this $\Teff$ regime, the NLTE effects on the Sc\ii\ lines are weak. From calculations of \citet{2018MNRAS.477.3343S}, we know that the same is true for lines of Ca\ii\ in the visual spectral range, which are widely used in the abundance determinations. For the hotter stars, using the LTE abundances may lead to a wrong classification of a star as Am. For example, \citet{2015PASP..127...58A} refer to HD~97633 ($\theta$~Leo) and HD~214994 ($o$~Peg) to as Am stars based on their subsolar abundances of Sc, supersolar abundances of Fe, and strongly enhanced heavy elements (Table~\ref{tab:cascsr} indicates abundances of the representative element Sr). However, for atmospheric parameters of $\theta$~Leo (9330/3.66) and $o$~Peg (9535/3.73), NLTE leads to the higher abundances of Sc, by 0.2 and 0.3~dex, respectively, removing completely a deficiency of Sc relative to the solar value. Similarly, the Ca abundances of these two stars increase in NLTE compared to LTE, by 0.07 and 0.23~dex, respectively, resulting in close-to-solar Ca/Sc abundance ratios. Thus, $\theta$~Leo and $o$~Peg are unlikely Am stars. These two stars are included in Table~\ref{tab:cascsr} for a discussion, but not in Fig.~\ref{fig:casc_lit}.

Table~\ref{tab:cascsr} includes a benchmark Am star HD~209625 (32~Aqr) from \citet{1997MNRAS.288..470A}, three Am stars from the Pleiades open cluster \citep{2008A&A...483..567G}, and six Am stars from the Hyades open cluster \citep{2010A&A...523A..71G}. Exceptions were made for HD~23631 and HD~27962 with $\Teff$ = 9610~K and 9025~K, respectively, because of their extremely large deficiency in Sc, which cannot be removed in NLTE. Table~\ref{tab:cascsr} lists abundances of Ca, Sc, Fe, and Sr, which are commonly used to separate Am from superficially normal stars. Absolute abundances from the cited papers were transformed to the [X/H] relative abundances using the solar system abundances from \citet{2021SSRv..217...44L}: $\eps{}$ = 6.27 (Ca), 3.04 (Sc), 7.45 (Fe), and 2.88 (Sr). As seen in Fig.~\ref{fig:casc_lit}, the Am stars reveal very similar Ca/Sc ratios, independent of the star's metallicity and effective temperature, with the mean [Ca/Sc] = 0.72$\pm$0.21. Thus, a highly supersolar Ca/Sc ratio, at the level of [Ca/Sc] $\simeq$ 0.6--0.7, can serve as a signature of the Am phenomenon.

The obtained results suggest a common mechanism for separation of calcium and scandium in the surface layers of Am stars. That mechanism is governed by the star's mass, mixing processes, mass loss \citep[][for scandium]{2022A&A...668A...6H} and, probably, by rotation, magnetic field, binarity, and other stellar parameters. Therefore, in atmospheres of different Am stars, it produces deviations of different magnitude from the solar Ca and Sc abundances. However, resulting relative Ca/Sc abundance is nearly constant in different Am stars and thus much less sensitive to stellar parameters. One can expect that a reproduction of our observational finding by the diffusion models would involve less number of poorly known free parameters and could fix the most important drivers of chemical separation.

The Sr/Sc abundance ratio is less useful for classifying a star as Am because of a big scatter of more than 1~dex for stars of close metallicity or effective temperature.

Based on our abundance comparisons, we checked the candidate chemically peculiar stars identified by \citet{2022AJ....164..255C} in the open cluster Stock~2. Of nine strongest candidate Am stars in their Table~4, the two stars, BD+59~431 and BD+59~428 have [Ca/Sc] = 0.64$\pm$0.19 and 0.63$\pm$0.19 and can firmly be classified as Am stars. Large abundance errors prevent making firm classifications of BD+58~393, BD+59~437, BD+59~418, and BD+58~423 with [Ca/Sc] = 0.44$\pm$0.40, 0.42$\pm$0.36, 0.23$\pm$0.36, and 0.22$\pm$0.23, respectively.
The stars BD+57 571, BD+58~478, and BD+59~438, with [Ca/Sc] = $-0.09\pm$0.32, 0.04$\pm$0.02, and 0.00$\pm$0.11, are very likely superficially normal A stars.

\section{Non-LTE abundance corrections for Sc\ii\ lines}\label{sect:nlte_corr}

\begin{table}
\centering
\caption{NLTE abundance corrections (dex) for lines of Sc\ii\ depending on effective temperature, surface gravity, and [Sc/H] abundance.}
\label{tab:dnlte}
	\begin{tabular}{crrrrrrr} %
	\hline\hline \noalign{\smallskip}
 $\lambda$ & \multicolumn{7}{c}{$\Teff$/1000} \\
\cline{2-8} 
 (\AA) & 7.0 & 8.0 & 8.5 &  9.0 &   9.5 &  10.0 &  11.0 \\
		\hline \noalign{\smallskip}
\multicolumn{8}{c}{$\logg$ =  4.0, [Sc/H] = 0} \\
 4246.82 & $-$0.05 & $-$0.08 & $-$0.06 &  0.02 &  0.23 &  0.46 &  0.63 \\
 4374.46 & $-$0.06 & $-$0.02 &  0.01 &  0.08 &  0.22 &  0.38 &  0.57 \\
 5526.80 & $-$0.06 & $-$0.01 &  0.02 &  0.08 &  0.21 &  0.37 &  0.57 \\
\multicolumn{8}{c}{$\logg$ =  4.0, [Sc/H] = $-0.5$} \\
 4246.82 & $-$0.05 & $-$0.04 &  0.02 &  0.14 &  0.32 &  0.47 &  0.61 \\
 4374.46 & $-$0.00 &  0.03 &  0.04 &  0.10 &  0.23 &  0.38 &  0.57 \\
 5526.80 &  0.02 &  0.03 &  0.04 &  0.09 &  0.22 &  0.37 &  0.57 \\
\multicolumn{8}{c}{$\logg$ =  4.0, [Sc/H] = $-1$} \\
 4246.82 &  0.00 &  0.07 &  0.12 &  0.19 &  0.31 &  0.45 &  \\
 4374.46 &  0.03 &  0.04 &  0.05 &  0.10 &  0.23 &  0.38 & \\
 5526.80 &  0.05 &  0.04 &  0.05 &  0.09 &  0.22 &  0.38 & \\
\multicolumn{8}{c}{$\logg$ =  1.5, [Sc/H] = 0} \\
 4246.82 &  0.00 &  0.13 &  0.49 &  0.91 &   &    &     \\
 4374.46 & $-$0.11 &  0.07 &  0.54 &  0.76 &   &    &     \\
 5526.80 & $-$0.22 &  0.03 &  0.46 &  0.64 &   &    &     \\
\noalign{\smallskip}\hline \noalign{\smallskip}
\end{tabular}

This table is available in its entirety, namely, for 10 lines of Sc\ii\ and $\logg$ = 3.5, 4.0, 4.5 for [Sc/H] = 0, $-0.5$, and $-1$, in a machine-readable form in the online journal. A portion is shown here for guidance regarding its form and content.
\end{table}

\begin{figure}
	\includegraphics[width=\columnwidth]{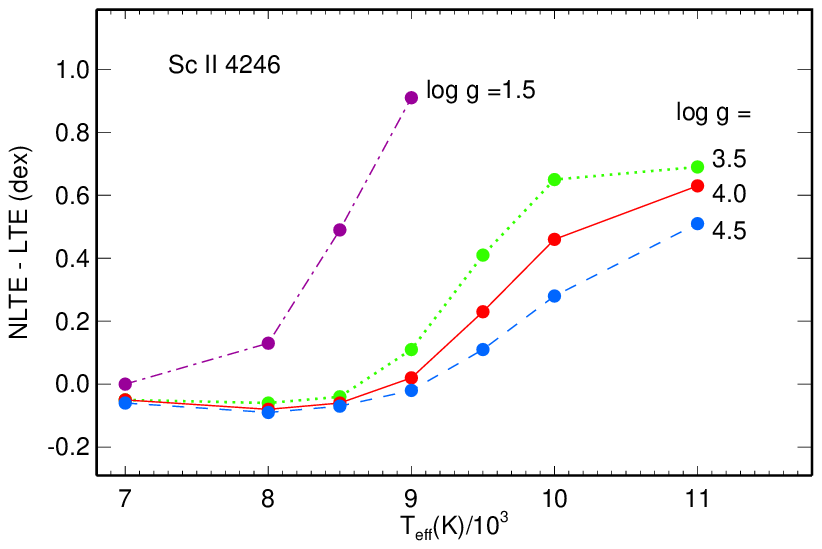}
	\includegraphics[width=\columnwidth]{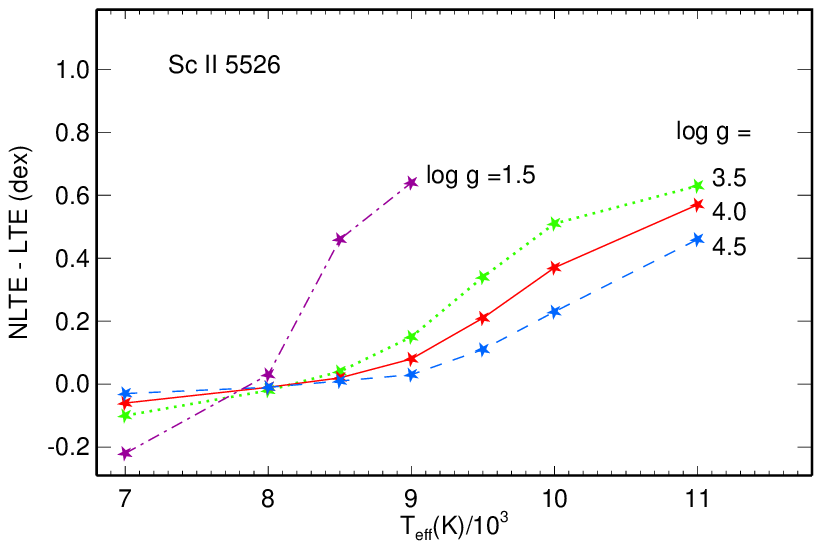}
    \caption{NLTE abundance corrections for Sc\ii\ 4246 (top panel) and 5526~\AA\ (bottom panel) depending on $\Teff$ in the model atmospheres with different $\logg$.
    Everywhere, [Sc/H] = 0.}
    \label{fig:dnlte}
\end{figure}

A grid of the NLTE abundance corrections for 10 lines of Sc\ii\ listed in Tables~\ref{tab:hd32115} and \ref{tab:ind_lines} was computed using the solar metallicity model atmospheres from the Kurucz's website\footnote{\url{http://kurucz.harvard.edu/grids/gridp00odfnew/}} (date: April 13 of 2011). Effective temperature ranges from 7000~K to 11\,000~K and $\logg$ takes values of 1.5 ($\Teff \le$ 9000~K), 3.5, 4.0, and 4.5.
Since A-B type stars may have rather different Sc abundances, the NLTE calculations were performed with [Sc/H] = 0, $-0.5$, and $-1.0$. The results are presented in Table~\ref{tab:dnlte}.

Figure~\ref{fig:dnlte} displays $\Delta_{\rm NLTE}$s for two selected lines from calculations with [Sc/H] = 0. The NLTE abundance corrections depend strongly on $\Teff$ and $\logg$. For $\Teff \le$ 9000~K and $\logg \ge$ 3.5, the NLTE effects are small because Sc\ii\ is the majority species in such atmospheres. For the hotter temperatures, an overionization of Sc\ii\ leads to weakened lines and positive abundance corrections, which grow to $\Delta_{\rm NLTE}$ = 0.5-0.6~dex for $\Teff$ = 10\,000-11\,000~K. As shown in Sect.~\ref{sect:abund},  the NLTE corrections of exactly such magnitude are needed to remove a deficiency of Sc in 21~Peg and $\pi$~Cet, obtained in the LTE analysis.

In the $\logg$ = 1.5 models, an overionization of Sc\ii\ shifts to the cooler temperatures and $\Delta_{\rm NLTE}$ grows dramatically with increasing $\Teff$. For example, $\Delta_{\rm NLTE}$ = 0.49 and 0.54~dex for Sc\ii\ 4246 and 4374~\AA, respectively, in the 8500/1.5 model. Exactly such great corrections are needed to obtain close-to-solar abundance of Sc for a prototype A-type supergiant Deneb. With $\Teff$ = 8525~K and $\logg$ = 1.10, \citet{2008A&A...479..849S} derived the LTE abundances $\eps{LTE}$ = 2.30 and 2.45 from Sc\ii\ 4246 and 4374~\AA, respectively. Using the recent $gf$-values, we transformed the LTE abundances to $\eps{LTE}$ = 2.34 and 2.27. Based on our NLTE calculations, we expect that the NLTE abundance of Sc in Deneb is $\eps{NLTE} >$ 2.9 and thus close to the solar one.


\section{Conclusions}\label{sect:Conclusions}

In this study, we developed the NLTE method for analyses of the scandium lines in A-B type stars and showed that NLTE is essential for determinations of stellar Sc abundances. The model atom of Sc\ii -Sc\iii\ was constructed for the first time. In the atmospheres with $\Teff \le$ 9000~K and $\logg \ge$ 3.5, Sc\ii\ is the majority species, and the NLTE abundance corrections for lines of Sc\ii\ do not exceed 0.1~dex in absolute value. In the hotter atmospheres, Sc\ii\ is subject to an overionization that leads to weakened lines and positive NLTE abundance corrections, which exceed 0.5~dex for $\Teff \ge$ 10\,000~K.

The NLTE abundances of Sc were determined for eight A9-B3 main-sequence stars with well-determined atmospheric parameters and high-quality observed spectra available. For each star with more than one scandium line measured, NLTE reduces the line-to-line scatter. We note, in particular, $\iota$~Her, for which the Sc abundance was derived from the four UV lines of Sc\iii\ with $\sigma$ = 0.04~dex and 0.17~dex in NLTE and LTE, respectively.

Our results provide one more evidence for a status of HD~32115, HD~73666, 21~Peg, $\pi$~Cet, and $\iota$~Her as superficially normal stars.
Their NLTE abundances (when subtracting the metallicity of the Praesepe open cluster for HD~73666), are consistent within the error bars with both the solar system Sc abundance, $\eps{met}$ = 3.04$\pm$0.03 \citep{2021SSRv..217...44L}, and the solar photosphere one, $\eps{\odot}$ = 3.12$\pm$0.05 \citep{nlte_sc2}.
NLTE is, in particular, important for 21~Peg and $\pi$~Cet, which reveal substantially subsolar LTE abundances of [Sc/H] = $-0.42$ and $-0.37$, respectively.
An extended sample of the superficially normal stars needs to be investigated with respect to their Sc abundances in order to fix the cosmic abundance standard for scandium.


NLTE reduces a deficiency of Sc in the atmospheres of our two Am stars, HD~72660 and Sirius, compared to the solar Sc abundance, however, it remains with [Sc/H] = $-0.18$ and $-0.80$, respectively.

Despite substantial discrepancies in the Sc abundance, different Am stars reveal very similar Ca/Sc ratios: [Ca/Sc] = 0.56$\pm$0.05, on average, for HD~180347 \citep{2023MNRAS.524.1044T}, HD~72660, and Sirius. We selected 13 Am stars with $\Teff \precsim$ 9000~K and the LTE abundances from the literature and obtained the mean [Ca/Sc] = 0.72$\pm$0.21. For comparison, [Ca/Sc] = $-0.01\pm$0.08 for our six superficially normal stars. In contrast to Ca/Sc, the Sr/Sc abundance ratios of the Am stellar sample reveal a big scatter of more than 1~dex for stars of close metallicity or effective temperature.
The obtained results suggest that peculiar abundances of Ca and Sc in the atmospheres of Am stars are produced by a common mechanism. In different Am stars, it can produce
deviations of different magnitude from the solar Ca and Sc abundances, however, holds nearly constant Ca/Sc abundance ratio. Different processes need to be involved in order to create enhancements in the heavy elements, as observed in Am stars.

Based on our empirical finding of a highly supersolar and nearly constant Ca/Sc abundance ratio of [Ca/Sc] = 0.6-0.7 in the Am stars, {\it we propose the Ca/Sc abundance ratio, but not abundances of individual Ca and Sc elements to be used for classifying a star as Am and for testing the diffusion models.}


We provide the NLTE abundance corrections for ten lines of Sc\ii\ in a grid of the solar-metallicity model atmospheres with $\Teff$ from 7000~k to 11\,000~K and $\logg$ = 1.5, 3.5, 4.0, and 4.5. For the lowest gravity, the NLTE effects are extremely strong, with $\Delta_{\rm NLTE}$ of up to 0.91~dex for Sc\ii\ 4246~\AA\ in the 9000/1.5 model. For $\logg$ = 3.5, 4.0, and 4.5, the NLTE calculations were performed with three different Sc abundances: [Sc/H] = 0, $-0.5$, and $-1.0$.

\section{Acknowledgments}
 This study made use of the ASTRAL, NIST, SIMBAD\footnote{\url{https://simbad.cds.unistra.fr/simbad/}}, VALD, ADS\footnote{\url{http://adsabs.harvard.edu/abstract\_service.html}}, and R.~Kurucz's databases.

\section{Data availability}

The data underlying this article will be shared on reasonable request to the corresponding author.

\bibliography{atomic_data,mashonkina,nlte,ab2023,si2018}
\bibliographystyle{mnras}

\label{lastpage}
\end{document}